\documentclass[a4paper, 12pt, oneside]{article}
\usepackage{amsmath}
\usepackage{amssymb}
\usepackage{enumerate}
\usepackage{graphicx}
\usepackage{epsf}
\usepackage{verbatim}
\usepackage{tikz}
\usepackage{systeme,mathtools}
\usetikzlibrary{matrix,arrows,decorations.pathmorphing}

\usepackage[amsmath,thmmarks]{ntheorem}
\newcounter{thm} 

\newtheorem{Thm}[thm]{Theorem}

\newtheorem{Lem}[thm]{Lemma}
\newtheorem{Prop}[thm]{Proposition}

\theoremstyle{nonumberplain}
\theoremheaderfont{\normalfont\itshape}
\theorembodyfont{\normalfont}
\theoremseparator{.}
\theoremsymbol{\ensuremath{\square}}
\newtheorem{proof}{Proof}

\def \Z {\mathbb Z}

\def \C {\mathbb C}
\def \P {\mathbb P}

\def \D {\mathcal{D}}

\def \M {\mathcal{M}}

\def \W {\mathcal{W}}

\def \d {\mathrm{d}}
\def \tr {\mathrm{Tr}}

\def \bpsi {\bar{\psi}}
\def \btau {\bar{\tau}}
\def \h {\hbar}

\begin{document}
\begin{titlepage}

\title{A connection between the Kontsevich-Witten and Brezin-Gross-Witten tau-functions}
\author{Gehao Wang}
\date{}
\maketitle

\begin{abstract}
The Brezin-Gross-Witten (BGW) model is one of the basic examples in the class of non-eigenvalue unitary matrix models. The generalized BGW tau-function $\tau_N$ was constructed from a one parametric deformation of the original BGW model using the generalized Kontsevich model representation. It is a tau-function of the KdV hierarchy for any value of $N\in\C$, where the case $N=0$ reduces to the original BGW tau-function. In this paper, we present a representation of $\tau_N$ in terms of the $W_{1+\infty}$ operators that preserves the KP integrability. This naturally establishes a connection between the (generalized) BGW and Kontsevich-Witten tau-functions using $GL(\infty)$ operators, both considered as the basic building blocks in the theory of matrix models and partition functions. 
\end{abstract}
\vspace{20pt}


\end{titlepage}
\section{Introduction}
In the work of ``M-Theory'' of matrix models, we have seen that partition functions of every matrix model can be constructed from elementary constituents, (see, e.g. \cite{ASMM0}, \cite{ASMM1}, \cite{ASMM2} and \cite{O}). The main basic constituent is the Kontsevich-Witten (KW) tau-function $Z_{KW}$, because various matrix models can be constructed from it. However, the decomposition of some complex matrix models involves not only $Z_{KW}$, but another important ingredient, which has been identified as the Brezin-Gross-Witten (BGW) model (\cite{AMM1}). The BGW model is defined as the unitary matrix integral (\cite{BG}, \cite{GEW}) 
\begin{equation*}
Z_{BGW}=\frac{1}{V_M}\int_{M\times M}[\d U]e^{\frac{1}{\hbar}\tr (J^{\dagger}U+UJ^{\dagger})},
\end{equation*}
with Haar measure $[\d U]$ and $V_M$ being the volume of unitary group. It can also be described in terms of the generalized Kontsevich matrix model as the following $M\times M$ Hermitian matrix integral \cite{MMS}
\begin{equation*}
Z_{BGW}=\frac{\int [\d\Phi] \exp\left(\tr (\frac{\Lambda^2\Phi}{\hbar}+\frac{1}{\hbar\Phi}-M\ln \Phi) \right)}{\int [\d\Phi] \exp\left(\tr (\frac{1}{\hbar\Phi}-M\ln \Phi)\right)},
\end{equation*}
where $\Lambda$ is the diagonal matrix. In fact, under the variables $q_{k}=\tr \Lambda^{-k}$, the partition function made from $Z_{BGW}$ in Kontsevich phase is also a tau-function for the KdV (2-reduced KP) hierarchy like $Z_{KW}$, (see e.g. \cite{GN} and \cite{GN2}), which means that both tau-functions do not depend on even times $q_{2k}$. This is one of the similar properties that $Z_{BGW}$ and $Z_{KW}$ have. The Virasoro constraints for $Z_{BGW}$ has been introduced in \cite{GN2}. The cut-and-join operator descriptions for $Z_{KW}$ and $Z_{BGW}$ were introduced in \cite{A1} and \cite{A} respectively. In \cite{NP} (see also \cite{NP1}), Norbury has constructed a new cohomology class, and conjectured that the tau-function $Z_{BGW}$ is a generating function of the so-called the $n$-point $\Theta$-class intersection numbers, (a recent proof is in \cite{CGG}). Besides, as indicated in \cite{NP}, $Z_{BGW}$ can be used as a building block to construct partition functions using the action of twisted loop group and translations, just like the role of $Z_{KW}$ in partition functions of cohomological field theories described by Givental in \cite{AG} and \cite{AG1}.

The generalized BGW model $Z_N$ was first introduced in \cite{MMS} as the following deformed generalized Kontsevich matrix model
\begin{equation*}
Z_{N}=\frac{\int [\d\Phi] \exp\left(\tr (\frac{\Lambda^2\Phi}{\h}+\frac{1}{\h\Phi}+(N-M)\ln \Phi) \right)}{\int [\d\Phi] \exp\left(\tr (\frac{1}{\h\Phi}+(N-M)\ln \Phi)\right)},
\end{equation*}
and its partition function $\tau_N$ is a tau-function of the MKP hierarchy  \cite{KMMM} with discrete time $N$. Other properties of $\tau_N$, including Virasoro constraints, cut-and-join operator description, quantum spectral curve and the fact that $\tau_N$ is a tau-function of KdV hierarchy for any $N\in\C$ are presented in \cite{A}. The matrix resolvent approach of $\tau_N$'s KdV tau-structure is given in \cite{DYZ}. In addition, $\tau_N$ can be interpreted as an isomonodromic tau-function \cite{BR}. 

The set of all tau-functions for the KP hierarchy forms an orbit under the action of the group $GL(\infty)$ constructed from the $\widehat{\mathfrak{gl}(\infty)}$ algebra (see, e.g., \cite{DJKM} and \cite{MJE}). This action is transitive. Therefore it guarantees the existence of $GL(\infty)$ operators that transform a tau-function (including the trivial one) into any other one. The $W_{1+\infty}$ algebra is an important sub-algebra of $\widehat{\mathfrak{gl}(\infty)}$. It is the central extension of the Lie algebra of differential operators on the circle. Under the boson-fermion correspondence, the bosonic $W_{1+\infty}$ operators provide a fundamental symmetry for the KP hierarchy \cite{FKN}. In this paper, we present the following $W_{1+\infty}$ representation of the generalized BGW tau-function.
\begin{Thm}\label{main}
The generalized BGW tau-function $\tau_N(q)$ can be expressed as 	
\begin{multline}\label{maineq}
	\tau_N(q)=\exp\left\{\frac{\h}{4}\widehat{M}_{-1}+\frac{\h}{4}\widehat{L}_{-1}+\frac{\h(1-4N^2)}{16} q_1 \right\} 
	 \exp\left\{\frac{\h}{4}\widehat{M}_{-1}+\frac{\h(2N+1)}{8}\widehat{L}_{-1}\right\} \\
	 \operatorname{OE}\left[-\frac{3}{4}\widehat{M}_{-1}-\frac{6N+3}{8}\widehat{L}_{-1}-\frac{\h}{8}\widehat{Q}_{-2}-\frac{\h(4N+1)}{16}\widehat{M}_{-2}- \frac{\h(4N^2-1)}{32}\left(\widehat{L}_{-2}-\frac{1}{2}q_2\right) \right]
	 \cdot 1,
\end{multline}
where the notation $\operatorname{OE}$ means the ordered exponential \footnote{We say $e^W$ is the ordered exponential of $A$, denoted by $\operatorname{OE}[A]=e^W$, if $\partial_{\h}e^W=Ae^W$ and the initial condition is $e^{W}(\h=0)=1$. The expression of $W$ can be computed from $A$ using the Magnus expansion \cite{M}. See Appendix A.1 for a brief review.} with the parameter $\h$. The operator $\widehat{L}_n$ is the Virasoro operator defined in Eq.\eqref{hatL},  $\widehat{M}_n$ is the cut-and-join type operator defined in Eq.\eqref{hatM} and $\widehat{Q}_n$ is the operator in $W^{(4)}$-algebra defined in Eq.\eqref{hatQ}.
\end{Thm}
When $N=0$, the formula \eqref{maineq} becomes a $W_{1+\infty}$ representation of the original BGW tau-function $\tau_{BGW}$. Our proof of Theorem \ref{main} is in Sect.\ref{S31}. The strategy starts from the basis vectors that generate the point of the Sato Grassmannian corresponding to $\tau_N$. Usually they can be used to express the bilinear fermionic representation of the tau-function, which is a Bogoliubov transformation of the vacuum. Here, our approach is to transform those basis vectors into a single differential operator acting on the basis of the trivial tau-function $1$, and then decompose this operator into several simple ones. The first exponential operator in Eq.\eqref{maineq} is a special case of the operator in the representation of the tau-function of Grothendieck's dessins d'enfants (see Eq.\eqref{dessins} below). Therefore, this formula may contain some combinatorial interpretations of both $\tau_N$ and $\tau_{BGW}$.

Since the KP integrability is preserved, relations between tau-functions using the $W_{1+\infty}$ operators are particularly interesting. For example, the Hurwitz tau-function has a well-known cut-and-join representation using the operator $\widehat{M}_0$, which has a natural combinatorial interpretation (\cite{GJ}). The tau-function of Grothendieck’s dessins d'enfants has a similar representation using the operators $\widehat{M}_{-1}$ and $\widehat{L}_{-1}$ (\cite{MZo}, \cite{PZ}). In \cite{MK}, Kazarian showed that the linear Hodge tau-function is indeed a tau-function for the KP hierarchy by connecting it to the Hurwitz tau-function using the Virasoro operators. Furthermore, it can also be connected to the Kontsevich-Witten tau-function using the Heisenberg-Virasoro operators (\cite{AE}, \cite{LW} and \cite{W}). These formulas can be used to study the properties of such tau-functions and their relations. For example, the Virasoro constraints for the linear Hodge tau-function were constructed in \cite{AE} and \cite{GW}. The equivalence relation among the Virasoro constraints, cut-and-join (evolution) equation and polynomial recursion relation for the linear Hodge integrals were established in \cite{GW}, using the formula in  \cite{LW}. The case of triple Hodge tau-function has been studied in the series of papers \cite{A4}, \cite{A5} and \cite{A6}. We summarize some connections among these tau-functions using the $W_{1+\infty}$ operators in the following diagram.
\begin{center}
	\begin{tikzpicture}[scale=5.5]
		\node (A) at (0,0) {\quad};
		\node (B) at (0.8,0) {Dessins d'enfants};
		\node (C) at (1.6,0) {\quad};
		\node (D) at (2.4,0) {KW};
		\node (E) at (0,0.5) {BGW};
		\node (F) at (0.8,0.5) {1};
		\node (G) at (1.6,0.5) {Hurwitz};
		\node (H) at (2.4,0.5) {Hodge};
		\path[->,font=\scriptsize,>=angle 90]
		(F) edge node[above]{Eq.\eqref{maineq}} (E)
		(F) edge node[left]{\cite{MZo}} (B)
		(F) edge node[above]{\cite{GJ}} (G)
		(G) edge node[above]{\cite{MK}} (H)
		(H) edge node[left]{\cite{AE} and \cite{LW}} (D);
	\end{tikzpicture}.
\end{center}

Recently, there is some great progress on the investigation of the BKP hierarchy and its relation to the KP(KdV) hierarchy, (see, e.g., \cite{A2}, \cite{A3}, \cite{HO}, \cite{MM1}, \cite{MMNO}, \cite{Or},  and references therein). In fact, the cut-and-join operator of $\tau_N$ in \cite{A} (although not considered as $\widehat{\mathfrak{gl}(\infty)}$ operator) turns out to be the BKP symmetry operator for $\tau_N$ (\cite{A2},\cite{A3}). In our context, we will focus on the framework of the KP hierarchy. 

The structure of this paper is as follows. In Sect.\ref{S2}, we give some materials about the infinite dimensional Lie algebra related to the KP hierarchy. Sect.\ref{S3} includes a brief review on the tau-functions and their relations appearing in the above diagram. In Sect.\ref{S34}, we also give some partial results similar to Theorem \ref{main} regarding the Kontsevich-Witten tau-function, and discuss the $W_{1+\infty}$ representation of the Kontsevich-Penner model.  

\section{The infinite dimensional Lie algebras}\label{S2}
In this section, we give a brief introduction on the elements of the infinite dimensional Lie algebras $\widehat{\mathfrak{gl}(\infty)}$ and $W_{1+\infty}$ when they act as the infinitesimal transformations on the solutions of the KP equations. For more details we refer the readers to \cite{AZ}, \cite{BTB}, \cite{DJKM}, \cite{Kac}, \cite{KacRR}, \cite{MJE} and other related materials. Note that, in order to avoid any confusion, we use the notation ``$\psi$'' to represent fermions and $\bpsi$ for the first chern class of the cotangent line bundle on the moduli space of stable maps. 
 
\subsection{Free fermions}\label{S21}

We first introduce the free fermions $\psi_m,\psi_n^{*}$ with $m,n\in\Z$ (the subsequent of fermions are integers),  satisfying the canonical anticommutation:
\begin{equation*}
[\psi_m,\psi_n]_+=[\psi_m^{*},\psi_n^{*}]_+=0, \quad [\psi_m,\psi_n^{*}]_+=\delta_{m,-n},
\end{equation*}
where $[X,Y]_+=XY+YX$. They generate the Clifford algebra $\mathcal{A}$, consisting of linear combinations of monomials
\begin{equation*}
\psi_{m_1}\dots \psi_{m_r}\psi_{n_1}^{*}\dots\psi_{n_s}^{*}, \quad\mbox{where } m_1<\dots<m_r \mbox{ and } n_1<\dots<n_s. 
\end{equation*}
We also divide the set of fermions into the following two classes
\begin{align*}
&\{\psi_m,\psi_n^* \} \mbox{ for } m\leq 0\mbox{ and } n<0 \mbox{ are called creation operators};\\
&\{\psi_m,\psi_n^* \} \mbox{ for } m> 0 \mbox{ and }n\geq 0 \mbox{ are called annihilation operators.}
\end{align*}
In other words, for the vacuum state $|0\rangle$ and dual vacuum state $\langle 0|$ , we have
\begin{align*}
&\psi_m |0\rangle=\psi_n^*|0\rangle=0 \quad \mbox{ for } m>0,n\geq 0\\
&\langle 0|\psi_m=\langle 0|\psi_n^*=0 \quad\mbox{ for } m\leq 0, n< 0.
\end{align*}

The Fermionic Fock space $F$ is generated by the basis vectors 
\begin{align*}
& \psi_{m_1}\dots \psi_{m_r}\psi_{n_1}^{*}\dots\psi_{n_s}^{*}|0\rangle \\
&\mbox{for } m_1<\dots,<m_r\leq 0 \mbox{ and } n_1<\dots<n_s< 0,
\end{align*}
and it can be decomposed into a direct sum of vector spaces $F=\bigoplus_{m\in\Z} F^{(m)}$ by introducing the charge of fermions. The charge of $\psi_n$ and $\psi_n^*$ are defined to be $1$ and $-1$ respectively. Then the charge of the element 
$$\psi_{m_1}\dots \psi_{m_r}\psi_{n_1}^{*}\dots\psi_{n_s}^{*}|0\rangle$$
is $r-s$. The space $F^{(m)}$ consists of all the states of charge $m$. The dual Fermionic Fock space $F^*$ is generated by the basis vectors
\begin{align*}
& \langle 0|\psi_{m_1}\dots \psi_{m_r}\psi_{n_1}^{*}\dots\psi_{n_s}^{*} \\
&\mbox{for } 0< m_1<\dots,<m_r \mbox{ and } 0\leq n_1<\dots<n_s.
\end{align*}

The shifted vacuum states $n|\rangle$ and $\langle n|$, sometimes called as the $n$-th charged vacuum states, are defined by
\begin{align*}
& |n\rangle
=
\begin{cases}
\psi_{-n+1}\dots \psi_0|0\rangle, & n>0 \\
\psi_{n}^*\dots \psi_{-1}^*|0\rangle, & n< 0 ;
\end{cases}\\
& \langle n|
=
\begin{cases}
\langle 0| \psi_{1}\dots  \psi_{-n},  & n<0 \\
\langle 0| \psi_0^*\dots \psi_{n-1}^*, & n>0.
\end{cases}.
\end{align*}
Therefore, by definition, we have
\begin{align*}
&\psi_m |n\rangle=0,\quad m> -n \quad\mbox{ and }\quad \psi_m^*|n\rangle=0, \quad m\geq n;\\
&\langle n|\psi_m=0,\quad m\leq -n \quad\mbox{ and }\quad \langle n|\psi_m^*=0, \quad m< n.
\end{align*}
We also define the totally empty vacuum state $|\infty\rangle$ as
\begin{equation*}
|\infty\rangle= \dots \psi_{-2}\psi_{-1}\psi_0|0\rangle \quad\mbox{and}\quad |-\infty\rangle= \dots \psi_{-3}^*\psi_{-2}^*\psi_{-1}^*|0\rangle.
\end{equation*}
This means that
\begin{equation*}\label{psiinf}
|0\rangle=  \psi_0^*\psi_1^*\psi_2^* \dots |\infty\rangle \quad\mbox{and}\quad |0\rangle=  \psi_1\psi_2\psi_3 \dots |-\infty\rangle.
\end{equation*}

The expectation value of fermions is defined as a pairing $F^*\times F \rightarrow \C$ under the following rules:
\begin{align*}
& \langle 1  \rangle=1;\quad\langle \psi_m  \rangle=\langle \psi^*_n  \rangle=\langle \psi_m\psi_n  \rangle=\langle \psi_m^*\psi_n^*  \rangle=0;\\
& \langle \psi_m\psi_n^*  \rangle=0, \quad\mbox{ if } n\geq 0;\\
& \langle \psi_m\psi_n^*  \rangle=\delta_{m,-n},\quad \mbox{ if } n< 0.
\end{align*}
For $a\in \mathcal{A}$, the expectation value of $a$ is usually denoted by $\langle a  \rangle$. If $w_1,w_2,\dots,w_r$ is a set of free fermions, then the expectation value $\langle w_1w_2\dots w_r  \rangle$ is given by Wick's theorem as follows,
\begin{align*}
&\langle w_1w_2\dots w_r  \rangle=0, \quad\mbox{ if $r$ is odd};\\
&\langle w_1w_2\dots w_r  \rangle=\sum_{\sigma}\mbox{sign}(\sigma)\langle w_{\sigma(1)}w_{\sigma(2)} \dots \langle w_{\sigma(r-1)}w_{\sigma(r)} \rangle, \quad\mbox{ if $r$ is even}.
\end{align*}
Here $\mbox{sign}(\sigma)$ means the sign of the permutation $\sigma$,  $\sigma(i)<\sigma(i+1)$ for $1\leq i\leq r-1$ and $\sigma(1)<\sigma(3)<\dots \sigma(r-1)$. In other words, the summation above runs all distinct pairings of $w_1,w_2,\dots,w_r$.

\subsection{The boson-fermion correspondence}\label{S22}
Let $B^{(m)}=\C[x_1,x_2,\dots;t,t^{-1}]t^m$ and $B=\bigoplus_{m\in\Z} B^{(m)}$. The space $B$ is known as the bosonic Fock space. The extra parameter $t$ is to identify the charges. The bosons $\alpha_n$ are known by the usual formulas
\[
\alpha_n=
\begin{cases}
-nx_{-n} & n<0 \\
\frac{\partial}{\partial x_n} &  n>0.
\end{cases}
\]
They form the Heisenberg algebra. On the other hand, for the two free fermionic fields
\begin{equation*}
\psi(z)=\sum_{n\in\Z}\psi_nz^{-n}\quad\mbox{ and }\quad \psi^{*}(z)=\sum_{n\in\Z}\psi_n^{*}z^{-n-1},
\end{equation*} 
the following equation
\begin{equation}\label{f0}
\alpha(z)=:\psi(z)\psi^{*}(z):=\sum_{n\in\Z}\alpha_nz^{-n-1}
\end{equation}
provides as a realization of bosons in terms of fermions. In fact, under this construction, we can see that
\begin{equation*}
\alpha_n=\sum_{j\in\Z}:\psi_{-j}\psi_{j+n}^*:\quad \mbox{ and }\quad [\alpha_m,\alpha_n]=m\delta_{m+n,0}
\end{equation*}
and $\alpha_0$ works as charge operator, such that the eigenvalues of $\alpha_0$ on fermions are exactly the charges. Here the normal product $:\psi_m\psi_n^*:$ of a quadratic monomial in Fermions is defined as
\begin{equation*}
:\psi_m\psi_n^*:
=\psi_m\psi_n^*-\langle \psi_m\psi_n^*\rangle=
\begin{cases}
\psi_m\psi_n^* & \mbox{if }  m\leq 0 \mbox{ or if } n> 0;\\
-\psi_n^*\psi_m & \mbox{if } m> 0 \mbox{ or if } n\leq 0.
\end{cases}
\end{equation*}

There exists a unique isomorphism $\sigma_m$ between the two Fock spaces $F^{(m)}$ and $B^{(m)}$, such that, for $|v\rangle\in F^{(m)}$ and $H(x)=\sum_{n=1}^{\infty}x_n\alpha_n$,
\begin{equation*}
\sigma_m\left(|v\rangle\right)= \langle m|e^{H(x)}|v\rangle .
\end{equation*}
Let $\sigma$ be the isomorphism between $F$ and $B$, such that the restriction map to $F^{(m)}$ is $\sigma_m$, and $\sigma |m\rangle=t^m$. This isomorphism can be considered as the first part of the boson-fermion correspondence, which shows us how to realize the Fermionic Fock space using bosonic Fock space. 

As the second part of the boson-fermion correspondence, we want to realize the action of the Fermions on the Fermionic Fock space in terms of the differential operators on the bosonic Fock space. We consider the field
\begin{equation*}
\phi(z)=\sum_{n\neq 0}\frac{\alpha_n}{-n}z^{-n}+\alpha_0\ln(z)+P,
\end{equation*}
so that $\alpha(z) =\partial_z \phi(z)$. The operator $P$ satisfies the commutation relation
\begin{equation*}
[\alpha_0,P]=1 \quad\mbox{ and }\quad [\alpha_n,P]=0 \mbox{ for }n\neq 0.
\end{equation*}
For $m\geq 1$, let $t=e^P$ and
\begin{equation*}
\Psi_{\pm m}(z)=:e^{\pm m\phi(z)}:= e^{\pm m\sum_{i=1}^{\infty}z^{i}x_i}e^{\mp m\sum_{i=1}^{\infty}\frac{z^{-i}}{i}\frac{\partial}{\partial x_i}}z^{\pm m\alpha_0}t^{\pm m}.
\end{equation*}
Here the normal ordering in bosonic Fock space means that $\alpha_n$ with $n>0$ and operator $P$ are moved to the right. We identify the free fermionic fields as
\begin{equation}\label{fff}
\psi(z) \rightarrow \Psi_1(z) \quad \mbox{ and }\quad \psi^*(z) \rightarrow \Psi_{-1}(z).
\end{equation} 
In general, for $m\geq 2$, 
\begin{align*}
&\partial^{m-1}\psi(z)\dots \partial\psi(z)\psi(z) \rightarrow \Psi_m(z),\\
&\partial^{m-1}\psi^*(z)\dots \partial\psi^*(z)\psi^*(z) \rightarrow \Psi_{-m}(z).
\end{align*}
This gives us the correspondence between Fermions in $F^{(m)}$ and differential operators on $B$. For $m=0$, the space $F^{(0)}$ can be generated by the field $:\psi(z)\psi^*(z):$, which is identified in Eq.\eqref{f0}.

\subsection{The $\widehat{\mathfrak{gl}(\infty)}$ algebra}\label{S23}

In this subsection, let us review the representations of the Lie algebras $\widehat{\mathfrak{gl}(\infty)}$ using the free fermions. Let $\mathfrak{gl}(\infty)$ be the Lie algebra of all infinite matrices $(a_{ij})_{i,j\in\Z}$, $a_{ij}\in\C$, with a finite number of non-zero entries. Let $E_{mn}=(\delta_{im}\delta_{jn})_{i,j\in\Z}$, $m,n\in\Z$ be the standard basis of $\mathfrak{gl}(\infty)$. We can construct the representation $r$ of $\mathfrak{gl}(\infty)$ as
\begin{equation*}
r(E_{mn})=\psi_{-m}\psi_n^{*}.
\end{equation*}
Note that 
\begin{equation*}
[\psi_{-m}\psi_n^*,\psi_{-m'}\psi_{n'}^*]=\delta_{n,m'}\psi_{-m}\psi_{n'}^*-\delta_{n',m}\psi_{-m'}\psi_{n}^*.
\end{equation*}
We can see that the commutation relations above agree with the elementary matrices $E_{mn}$. Let $\widetilde{\mathfrak{gl}}(\infty)$ be the Lie algebra of all infinite matrices $(a_{ij})_{i,j\in\Z}$ with the condition: 
\begin{equation*}
\mbox{ there exists } k>0 \mbox{ such that } a_{ij}=0 \mbox{ for all } |i-j|>k. 
\end{equation*}
This means that matrices in $\widetilde{\mathfrak{gl}}(\infty)$ have a finite number of non-zero diagonals, and they can be expressed as finite linear combinations of the matrices in the form
\begin{equation*}
\sum_{i\in\Z} \lambda_iE_{i,i+n}
\end{equation*}
for some coefficients $\lambda_i\in\C,n\in\Z$. Sometimes they are called band matrices. Although the algebra $\widetilde{\mathfrak{gl}}(\infty)$ contains $\mathfrak{gl}(\infty)$, the representation $r$ can not extend to $\widetilde{\mathfrak{gl}}(\infty)$. For example, since $[\psi_{-i},\psi_{i}^*]_{+}=1$, the state
$$r\left(\sum_{i\in \Z}E_{ii}\right)|0\rangle=\sum_{i\in\Z}\psi_{-i}\psi_{i}^*|0\rangle$$ 
is not defined. Hence we need to consider the following projective representation
\begin{equation}\label{prep}
\widehat{r}(E_{mn})=:\psi_{-m}\psi_n^{*}:.
\end{equation}
Then, for two matrices $A$ and $B$ in $\widetilde{\mathfrak{gl}}(\infty)$,
\begin{equation*}
[\widehat{r}(A),\widehat{r}(B)]=\widehat{r}([A,B])+\tr([J,A]B)I,\quad J=\sum_{i\leq 0}E_{ii},
\end{equation*}
where $I$ represents the identity element. We define the Lie algebra $\widehat{\mathfrak{gl}(\infty)}$ to be the central extension $\widetilde{\mathfrak{gl}}(\infty)+\C K$, where $K$ is the central element with $\widehat{r}(K)=I$.

The boson-fermion correspondence provides us a way to realize the $\widehat{\mathfrak{gl}(\infty)}$ elements as differential operators on the space of functions. For $|z|>|w|$, taking into account Eq.\eqref{fff}, we consider the formal distribution
\begin{equation*}
:\psi(z)\psi^*(w):=\sum_{i,j\in\Z} \widehat{r}(E_{ij}) z^{i}w^{-j-1},
\end{equation*}
and the series 
\begin{align*}
Z(z,w)&=\sum_{i,j\in\Z} Z_{ij}z^{i}w^{-j-1}\nonumber\\ &=\frac{z^mw^{-m}}{z-w}\exp\left(\sum_{j=1}^{\infty}(z^j-w^j)x_j \right) \exp\left(- \sum_{j=1}^{\infty}\frac{z^{-j}-w^{-j}}{j}\frac{\partial}{\partial x_j}\right)-\frac{1}{z-w}.
\end{align*}
Then, the correspondence $\widehat{r}(E_{ij}) \rightarrow Z_{ij}$ determines a representation of $\widehat{\mathfrak{gl}(\infty)}$ elements as differential operators on the bosonic Fock space $B^{(m)}$. 

\subsection{The $W_{1+\infty}$ algebra}\label{S24}
Consider the Lie algebra $\D$ of differential operators on the circle
\begin{equation*}
z^n\partial_z^m, \mbox{ where } n,m\in\Z,m\geq 0.
\end{equation*}
We can take the operators
\begin{equation*}
J_n^k=z^{n+k}(-\partial_z)^k, \mbox{ where } n,k\in\Z,k> 0,
\end{equation*}
to form a basis of $\D$, and have the following embedding $\varphi$ of $\D$ into $\widetilde{\mathfrak{gl}}(\infty)$
\begin{equation*}
\varphi\left(\frac{1}{k!}J_n^k\right)=(-1)^k\sum_{j\in\Z}\binom{-j}{k}E_{j-n,j}.
\end{equation*}
Let $D=z\partial_z$. The operators $z^nD^k$ with $n,k\in\Z$ and $k\geq 0$ also form a basis for the algebra $\D$. For any two polynomials $f$ and $g$, we have the commutation relation
\begin{equation}\label{Dcomm}
[z^nf(D),z^mg(D)]=z^{n+m}f(D+m)g(D)-z^{n+m}f(D)g(D+n).
\end{equation}
Let $W_{1+\infty}=\D+\C C$ be the central extension of $\D$ defined by the cocycle restricted to $\varphi(\D)$. We can extend $\varphi$ into the homomorphism
\begin{equation*}
\widehat{\varphi}: W_{1+\infty} \rightarrow \widehat{\mathfrak{gl}(\infty)} \quad\quad\mbox{ with } \widehat{\varphi}(C)=K.
\end{equation*}

The $W_{1+\infty}$ algebra is known to have a free fermion realization. For the representation $\widehat{r}$ in Eq.\eqref{prep}, suppose $a$ is an operator in $\D$, then we can realize $W_a$ as
\begin{equation*}
\widehat{r}\left(\widehat{\varphi}(W_a)\right)=\text{res}_z\left(:\psi^{*}(z)a\psi(z):\right).
\end{equation*}
In the rest of our context, we omit the representation map, and write 
\begin{equation}\label{FR}
W_a=\text{res}_z\left(:\psi^{*}(z)a\psi(z):\right)
\end{equation}
for simplicity. On the other hand, as introduced in \cite{FKN}, for $m\geq 0, m\in\Z$, the following field
\begin{align}
W^{(m+1)}(z)&=\sum_{n\in\Z}W_n^{(m+1)}z^{-n-m-1}\nonumber\\
&=\sum_{l=0}^m\frac{(-1)^{l}}{(m+1-l)l!}\left(\frac{m!}{(m-l)!}\right)^2\frac{(2m-l)!}{(2m)!}\partial_z^{l}P^{(m+1-l)}(z) \label{field}
\end{align}
is a bosonic realization and it corresponds to the standard basis of the $W_{1+\infty}$ algebra, where
\begin{equation*}
	P^{(i)}(z)=:e^{-\phi(z)}\partial_z^ie^{\phi(z)}:.
\end{equation*}
For example, we have (see, e.g., \cite{AE} and \cite{FKN})
\begin{equation*}
W^{(1)}(z)=\sum_{n\in\Z}\alpha_nz^{-n-1}\quad\mbox{ and }\quad
W^{(2)}(z)=\sum_{n\in\Z}L_nz^{-n-2},
\end{equation*}
where the nodes 
\begin{equation}\label{nodes1}
\alpha_n=W_{-z^n} \quad\mbox{ and }\quad L_n=W_{-z^{1+n}\partial_z-\frac{n+1}{2}z^n}
\end{equation}
span the Heisenberg and Virasoro algebra respectively. Furthermore, we have
\begin{equation*}
W^{(3)}(z)=\sum_{n\in\Z}M_nz^{-n-3},
\quad\mbox{ and }\quad W^{(4)}(z)=\sum_{n\in\Z}Q_nz^{-n-4},
\end{equation*}
where the nodes $M_n$ and $Q_n$ are in the form
\begin{align}
&M_n=W_{-z^{2+n}\partial^2_z-(n+2)z^{1+n}\partial_z-\frac{z^n}{6}(n+1)(n+2)}.\label{nodes2}\\
&Q_n=W_{-z^{n+3}\partial_z^3-\frac{3}{2}(n+3)z^{n+2}\partial_z^2-\frac{3}{5}(n+2)(n+3)z^{n+1}\partial_z-\frac{1}{20}(n+1)(n+2)(n+3)z^n}\label{nodes3}.
\end{align}
In terms of the differential operators on bosonic Fock space in variables $q_k=kx_k$, we write the bosonic $W_{1+\infty}$ operators corresponding to $W_a$ as $\widehat{W}_a$. In particular, the Virasoro operators corresponding to $L_n$ as
\begin{align}\label{hatL}
\widehat{L}_n&=\frac{1}{2}\sum_{a+b=n}:\alpha_a\alpha_b:\nonumber\\
&=\sum_{i>0,i+n>0} (i+n)q_i\frac{\partial}{\partial q_{i+n}}+\frac{1}{2}\sum_{i+j=n} ij\frac{\partial^2}{\partial q_i\partial q_j}+\frac{1}{2}\sum_{i,j>0,i+j=-n}q_iq_j.
\end{align}
For the operators $M_n$ belonging to the $W^{(3)}$-algebra, the operators $\widehat{M}_n$  are expressed as
\begin{align}\label{hatM}
\widehat{M}_n&=\frac{1}{3}\sum_{a+b+c=n}:\alpha_a\alpha_b\alpha_c:\nonumber\\
&=\sum_{\substack{i,j>0\\i+j+n>0}}(i+j+n)q_iq_j\frac{\partial}{\partial q_{i+j+n}}+\sum_{\substack{i,j>0\\i+j-n>0}}ijq_{i+j-n}\frac{\partial^2}{\partial q_i\partial q_j}\nonumber\\
&\quad\quad+\frac{1}{3}\sum_{\substack{i,j>0\\n-i-j>0}}ij(n-i-j)\frac{\partial^3}{\partial q_i\partial q_j\partial q_{n-i-j}}
+\frac{1}{3}\sum_{\substack{i,j>0\\-n-i-j>0}}q_iq_jq_{-n-i-j}.
\end{align}
The operators $\widehat{Q}_n$ are 
\begin{align}\label{hatQ}
	\widehat{Q}_n&=\frac{1}{4}\sum_{a+b+c+d=n}:\alpha_a\alpha_b\alpha_c\alpha_d:-\frac{1}{4}\sum_{a+b=n}(a+1)(b+1):\alpha_a\alpha_b:\nonumber\\
	&\quad\quad+\frac{1}{10}(n+2)(n+3)\widehat{L}_n.
\end{align}
In particular,
\begin{align}\label{Q2}
	\widehat{Q}_{-2}&=\sum_{i,j,k>0}(i+j+k-2)q_iq_jq_k\frac{\partial}{\partial q_{i+j+k-2}}+\sum_{i,j,k>0}ijkq_{i+j+k+2}\frac{\partial^3}{\partial q_i\partial q_j \partial q_k}\nonumber\\
	&+\frac{3}{2}\sum_{\substack{i,j,k>0\\i+j-k>2}}k(i+j-2-k)q_iq_j\frac{\partial^2}{\partial q_k\partial q_{i+j-2-k}}+\frac{1}{2}\sum_{i>0}i(i+1)^2q_{i+2}\frac{\partial}{\partial q_i}.
\end{align}
In general, Eq.\eqref{FR} and \eqref{field} can be used to determine the operators $\widehat{W}_a$ for any $a\in\D$.

\section{The tau-functions of KP hierarchy}\label{S3}
The tau-functions $\tau$ of the KP hierarchy in fermionic representations lie in $F^{(0)}$. They form the orbit of the vacuum state $|0\rangle$ under the action of the group $GL(\infty)$ associated with the Lie algebra $\widehat{\mathfrak{gl}(\infty)}$. The group elements are defined as
\begin{equation*}
e^{X_1}e^{X_2}\dots e^{X_k}
\end{equation*}
with $X_1,X_2,\dots,X_k\in\widehat{\mathfrak{gl}(\infty)}$, that is, their corresponding matrices are band matrices (\cite{MJD}). Suppose $g$ is an element in $GL(\infty)$. Since
\begin{equation*}
|0\rangle=  \psi_0^*\psi_1^*\psi_2^* \dots |\infty\rangle,
\end{equation*}
and $g|\infty\rangle=|\infty\rangle$, the action of $g$ on the vacuum state $|0\rangle$ can be described as
\begin{equation*}
g|0\rangle=  g\psi_0^*g^{-1}\cdot g\psi_1^*g^{-1} \dots |\infty\rangle,
\end{equation*} 
where each $g\psi_i^*g^{-1}$ is a linear combinations of fermionic operators. The expectation value
\begin{equation*}
\tau=\frac{\langle 0|e^{H(x)} g|0\rangle}{\langle 0| g|0\rangle}
\end{equation*} 
obeys Hirota bilinear equations, and it defines a tau-function for the KP hierarchy. 

In the language of the infinite dimensional Grassmannian, the tau-function $\tau$ can be written as the determinant
\begin{equation}\label{det}
\tau=\det\left[ \oint \frac{\mathrm{d} t}{2\pi i}\exp(\sum_{k\geq 1} z^kx_k)z^{-m}\Phi_n(z)\right]_{m,n\geq 1},
\end{equation}
for some functions $\Phi_n(z)$ that make the determinant well-defined. These functions $\{\Phi_n(z)\}$ are referred as the admissible basis vectors for the tau-function $\tau$. They define a space $\mathcal{W}$ which is a point on the big cell of the Sato Grassmannian \cite{S}, and can be transformed into the following form
\begin{equation}\label{Phi}
\Phi_j(z)=z^{j-1}+\sum_{i>1-j} \phi_{j,i}z^{-i},\quad j\geq 1.
\end{equation}
For each $\mathcal{W}$, we can further transform an admissible basis into a unique basis with $\phi_{j,i}=0$ for $i<1$, which is usually called the canonical basis (or affine coordinates) for the tau-function $\tau$. If we go back to the fermionic language, we can find a matrix $(B_{mn})$, such that, for
\begin{equation}\label{G}
G=\exp\left(\sum_{m,n\in\Z}B_{m,n}:\psi_{m}\psi_{n}^*:\right),
\end{equation}
we have
\begin{equation}\label{phi}
\widetilde{\phi}_{j-1}=G\psi_{j-1}^*G^{-1}=\psi_{j-1}^*+\sum_{i>1-j} \phi_{j,i}\psi_{-i}^*=\text{res}_z\left(\Phi_j(z)\psi^{*}(z)\right),
\end{equation}
and $G|0\rangle$ corresponds to the tau-function $\tau$. There exists an element $\widetilde{G}$ in the form
\begin{equation}\label{tildeG}
\widetilde{G}=\exp\left(\sum_{m\geq 0,n\geq 1}A_{m,n}\psi_{-m}\psi_{-n}^*\right),
\end{equation}
such that
\begin{equation*}
 G|0\rangle=\langle 0| G|0\rangle \times \widetilde{G}|0\rangle.
\end{equation*}
Therefore, the tau-function $\tau$ in fermionic representation can be written as 
\begin{equation*}
\tau=\widetilde{G} |0\rangle= \widetilde{\psi}_{0}\widetilde{\psi}_{1}\widetilde{\psi}_{2}\dots  |\infty\rangle,
\end{equation*}
where, for $m=0,1,2,\dots$,
\begin{equation*}
\widetilde{\psi}_{m} =\widetilde{G}\psi_{m}^*\widetilde{G}^{-1}
=\psi_{m}^*-\sum_{n=1}^{\infty} A_{m,n}\psi_{-n}^*.
\end{equation*}
The set of fermionic operators $\{\widetilde{\psi}_{n}\}$ corresponds to the canonical basis vectors $\{\Psi_{m+1}(z)\}$ for the tau-function $\tau$, where
\begin{equation*}
\Psi_{m+1}(z)=\langle 0|\psi_{m}^*\psi(z)\widetilde{G}|0\rangle=z^m-\sum_{n=1}^{\infty} A_{m,n}z^{-n}.
\end{equation*}
Using the boson-fermion correspondence and Wick's theorem, we can write the tau-function $\tau$ in the form of correlation function
\begin{equation*}
\tau\left(-[Z^{-1}]\right)=\frac{\det_{i,j=1}^M\Psi_j(z_i)}{\Delta(z)}
\end{equation*}
for some integer $M$, where $\Delta(z)=\prod_{i<j}(z_j-z_i)$ is the Vandermonde determinant, (the basis vectors in the above equation can be replaced  by $\Phi_j$). In fact, we can see that the set of vectors $\{z^m\}, (m\geq 0),$ form a basis for the trivial tau-function $1$. Here the notation $\tau\left(-[Z^{-1}]\right)$ means that our tau-function is in the following Miwa parametrization for the matrix model,
\begin{equation*}
\tau\left(-[Z^{-1}]\right)=\tau\left(x_k=\frac{q_{k}}{k}=-\frac{1}{k}\tr \Lambda^{-k}\right).
\end{equation*} 

For an operator $a$ in $\D$, the action of $\exp\left(W_{a}\right)$ on the tau-function $\tau$ is equivalent to the action of $e^a$ on the basis vectors $\Psi_{m+1}(z)$ (or  $\Phi_{m+1}(z)$), (see, e.g, \cite{FKN} and \cite{AE}). This can be demonstrated as the following. Suppose $a^*\in\D$ is the adjoint operator of $a$ satisfying 
\begin{equation}\label{a*}
	\text{res}_z\left(f(z)(a\cdot g(z))\right)=\text{res}_z\left(g(z)(a^*\cdot f(z))\right)
\end{equation}
for any functions $f(z)$ and $g(z)$. Then,
\begin{equation*}
	\text{res}_z\left(f(z)(e^a\cdot g(z))\right)=\text{res}_z\left(g(z)(e^{a^*}\cdot f(z))\right).
\end{equation*}
By the definition Eq.\eqref{FR}, we have
\begin{equation*}
[W_a,\psi^*(z)]={a^*}\cdot \psi^*(z).
\end{equation*}
Therefore,
\begin{align}\label{eq}
e^{W_{a}}\widetilde{\psi}_{m}e^{-W_{a}}&=\text{res}_z\left(\Psi_{m+1}(z)e^{W_{a}}\psi^{*}(z)e^{-W_{a}}\right)\nonumber\\
&=\text{res}_z\left(\Psi_{m+1}(z)e^{a^*}\cdot \psi^*(z)\right)\nonumber\\
&=
\text{res}_z\left(\psi^{*}(z)e^a\cdot\Psi_{m+1}(z)\right).
\end{align}
In terms of the Sato Grassmannian, the action of $e^a$ on the space $\mathcal{W}_G$ determined by the group element $G$ will result in the space $\mathcal{W}_{e^{W_{a}}G}$ determined by the group element $e^{W_{a}}G$. Then, in bosonic Fock space, the relation between the corresponding tau-functions can be described by the following equation
\begin{equation}\label{WG}
\tau_{e^{W_{a}}G}(q)=\frac{\langle 0|G|0\rangle}{\langle 0|e^{W_{a}}G|0\rangle}e^{\widehat{W_{a}}}\tau_G(q)=\frac{1}{\langle 0|e^{W_{a}}\widetilde{G}|0\rangle}e^{\widehat{W_{a}}}\tau_G(q),
\end{equation}
where we let $\tau(q)$ be the tau-function $\tau$ in variables $q_i$. If $W_a$ consists of fermions $\psi_m\psi_n^*$ with positive energy \footnote{In this context, the energy of fermions $\psi_m$ and $\psi_n^*$ are defined to be $\frac{1}{2}-m$ and $-\frac{1}{2}-n$ respectively. So the energy of $\psi_m\psi_n^*$ is $-(m+n)$.}, then the constant $\langle 0|e^{W_{a}}\widetilde{G}|0\rangle$ equals to $1$. 

If the operator $a$ satisfies $a\W\subset \W$, then for the tau-function $\tau$ corresponding to $\W$, we have $\widehat{W_{a}}\cdot \tau=C\tau$ for some constant $C$. Such operator $a$ is usually called the Kac-Schwarz operator for the tau-function $\tau$ \cite{KacS}.

In our context, we will also use another way of Miwa parametrization, that is,
\begin{equation*}
x_k=\frac{q_{k}}{k}=\frac{1}{k}\tr \Lambda^{-k}.
\end{equation*}
 Under this parametrization, for the same group element $\widetilde{G}$ in Eq.\eqref{tildeG}, the canonical basis vectors $\{\Psi^{\perp}_{m+1}(z)\}$ for the tau-function $\tau\left([Z^{-1}]\right)$ are in the form 
\begin{equation}
\Psi^{\perp}_{m+1}(z) =\langle 0|\psi_{m+1}\psi^*(z)\widetilde{G}|0\rangle=z^{m}+\sum_{k=1}^{\infty} A_{k-1,m+1}z^{-k}.
\end{equation}
They are usually referred as the adjoint canonical basis vectors and define the orthogonal complement $\mathcal{W}^{\perp}$ of the subspace $\mathcal{W}$, since
\begin{equation}\label{or}
\text{res}_z\left(\Psi_{i}(z)\Psi^{\perp}_{j}(z)\right)=0,  \mbox{ for }i,j\geq 1.
\end{equation}
For $a$ in $\D$,  we can find the corresponding operator $W_{a'}$ in $W_{1+\infty}$ with $a'\in \D$, such that the action of $e^a$ on $\Psi^{\perp}_{m+1}(z)$ is equivalent to the action of $\exp\left(W_{a'}\right)$ on the tau-function. 
Let $\sum c_{j}e^{a}\cdot \Psi^{\perp}_{j}(z)$ be an adjoint canonical basis vector for some coefficients $c_j$. Then, by Eq.\eqref{or}, we need to find an operator $a'$, such that any canonical basis vector in the form $\sum d_{i}e^{a'}\cdot \Psi_{i}(z)$ satisfies the following equation
\begin{equation}\label{a'a}
\text{res}_z\left((\sum d_{i}e^{a'}\cdot \Psi_{i}(z))(\sum c_{j}e^{a}\cdot \Psi^{\perp}_{j}(z))\right)=0.
\end{equation}
Using the adjoint operator $a^*$, we have
\begin{align*}
&\text{res}_z\left( (e^{a'}\cdot \Psi_{i}(z))(e^{a}\cdot \Psi^{\perp}_{j}(z))\right)\\
=&\text{res}_z\left( \Psi^{\perp}_{j}(z)(e^{a^*}e^{a'}\cdot \Psi_{i}(z) )\right).
\end{align*}
Therefore, by Eq.\eqref{or}, we have obtained the choice $a'=-a^*$. In fact, for several operators $\{a_1,a_2,\dots,a_n\}$ in $\D$, we have
\begin{equation*}
	\text{res}_z\left(f(z)(e^{a_1}e^{a_2}\dots e^{a_n}\cdot g(z))\right)
	=\text{res}_z\left((e^{a_n^*}\dots e^{a_2^*}e^{a_1^*}\cdot f(z)) g(z)\right).
\end{equation*}
Therefore,
\begin{align}\label{several}
	&\text{res}_z\left( (e^{-a_1^*}\dots e^{-a_n^*}\cdot\Psi_{i}(z))(e^{a_1}\dots e^{a_n}\cdot  \Psi^{\perp}_{j}(z))\right)\nonumber\\
	=&\text{res}_z\left( \Psi^{\perp}_{j}(z)\Psi_{i}(z) \right).
\end{align}
Here we state some examples of the adjoint operators in $\D$,
\begin{equation}\label{adex}
	(z^n)^*=z^n,  \quad (z^n\partial_z^k)^*=(-\partial_z)^kz^n,\quad \mbox{and}\quad(z^nD^k)^*=z^{-1}(-D)^kz^{n+1}.
\end{equation}
The above argument also holds for the basis vectors $\Phi_{i}$.

\subsection{The generalized BGW tau-function $\tau_N$}\label{S31}

From the results in \cite{MMS} and \cite{A}, we know that the determinantal representation of $Z_{N}$ for large values of $\lambda$ is 
\begin{equation*}
Z_{N}(\Lambda)=\frac{\det_{i,j=1}^M\Phi^{(N)}_j(\lambda_i)}{\Delta(\lambda)},
\end{equation*}
where the basis vectors $\Phi_j^{(N)}(z)$ for $Z_N$ are in the form  
\begin{equation}\label{Phi1}
	\Phi_j^{(N)}(z)=z^{j-1}+z^{j-1}\sum_{k=1}^{\infty}\frac{(-\h z^{-1})^k}{4^kk!}\prod_{s=1}^{k}\left((j-1-N)^2-(s-\frac{1}{2})^2\right).
\end{equation}
In fact, all the basis vectors can be generated by $\Phi_1^{(N)}(z)$ using a Kac-Schwarz operator $K_a$ as follows,
\begin{multline}\label{Phij+1}
\Phi_{j+1}^{(N)}(z)=\left(K_a-\frac{1}{4}+\frac{N}{2}-j+1\right)\left(K_a-\frac{1}{4}+\frac{N}{2}-j+2\right)\\
\dots\left(K_a-\frac{1}{4}+\frac{N}{2}\right)\cdot \Phi_1^{(N)}(z),
\end{multline}
where $K_a=z\h^{-1}+\frac{1}{2}D$. They define a point on the big cell of the Sato Grassmannian 
\begin{equation}\label{WN}
\mathcal{W}^{\perp}_N=\langle \Phi_1^{(N)}(z),\Phi_2^{(N)}(z),\dots\rangle,
\end{equation}
which is the orthogonal complement of the space $\mathcal{W}_N$ corresponding to the generalized BGW tau-function $\tau_{N}$. In fact, $\tau_N$ is a series in $N^2$. The case $N=\frac{1}{2}$ gives us the trivial tau-function. When $N=l+\frac{1}{2}$ for $l=1,2,3,\dots$, the tau-function $\tau_N$ is the Schur function corresponding to the partition $(l,l-1,\dots,1)$, (see Theorem 3.9 in \cite{A}). 

\vspace{10pt}
\noindent
{\bf Remark:} 
\emph{
Let us consider the following operator
\begin{equation}\label{M}
\M_j= z^{-1}\left((j-1-N)^2-(D-\frac{1}{2})^2\right).
\end{equation}
Observe that
\begin{equation*}
\left((j-1-N)^2-(D-\frac{1}{2})^2\right)\cdot z^{-k}=\left((j-1-N)^2-(k+\frac{1}{2})^2\right)z^{-k}.
\end{equation*}
From, Eq.\eqref{Phi1}, we can see that the basis vectors $\Phi_{j}^{(N)}(z)$ can be written as
\begin{equation}\label{phij}
\Phi_{j}^{(N)}(z)=z^{j-1}e^{-\frac{\h}{4}\M_j}\cdot 1.
\end{equation}
The operator $\M_1$ (also appeared in \cite{A2}) will be used in our later proofs. Using Eq.\eqref{adex}, the adjoint operator $\M_1^*$ is in the form
\begin{equation*}
	\M_1^*=z^{-1}\left(N^2-(D+\frac{1}{2})^2\right)=-z^{-1}D^2-z^{-1}D+\left(N^2-\frac{1}{4}\right)z^{-1}.
\end{equation*}
From Eq.\eqref{nodes1} and \eqref{nodes2}, we have $W_{-z^{-1}D}=L_{-1}$ and  $W_{-z^{-1}D^2}=M_{-1}$. Then,
\begin{align}\label{M1}
\M_1^*=M_{-1}+L_{-1}+\left(\frac{1}{4}-N^2\right)\alpha_{-1}.
\end{align}
It corresponds to the operator in Eq.\eqref{tauNM1}.
}

In the rest of this subsection, we will prove Theorem \ref{main}. Here we define the following operator-valued function and its inverse to help with our proofs in the next two lemmas. Namely, for arbitrary $N$, 
\begin{align}
	&\Gamma(D-N+\frac{1}{2})\cdot z^k=\Gamma(k-N+\frac{1}{2})z^k,\label{gammao} \\
	& \frac{1}{\Gamma(D-N+\frac{1}{2})}\cdot z^k=\frac{1}{\Gamma(k-N+\frac{1}{2})}z^k\nonumber.
\end{align}
In this way, for $n\geq 1, n\in\Z$, we can express the operator 
\begin{equation*}
	\left(z^{-1}(D-N-\frac{1}{2})\right)^n=z^{-n}\prod_{i=0}^{n-1}\left(D-N-\frac{1}{2}-i\right)
\end{equation*}
as the conjugation
\begin{align}
	  \frac{1}{\Gamma\left(D-N+\frac{1}{2}\right)}z^{-n}\Gamma\left(D-N+\frac{1}{2}\right)=&\left\{\frac{1}{\Gamma\left(D-N+\frac{1}{2}\right)}z^{-1}\Gamma\left(D-N+\frac{1}{2}\right) \right\}^n\nonumber\\
	  =&\left(z^{-1}(D-N-\frac{1}{2})\right)^n\nonumber\\
	  =&z^{-n}\frac{\Gamma\left(D-N+\frac{1}{2}\right)}{\Gamma\left(D-N-n+\frac{1}{2}\right)}.\label{gammaconj}
\end{align}

\vspace{6pt}
\noindent
{\bf Remark}: 
\emph{ One can realize the operator $\Gamma(D-N+\frac{1}{2})$ as, (see, e.g.,\cite{A3})
	\begin{equation*}
		\Gamma\left(D-N+\frac{1}{2}\right)=\exp\left\{ (D-N)\log(N)-N+\frac{1}{2}\log(2\pi)-\sum_{n=1}^{\infty}\frac{B_{n+1}(D+\frac{1}{2})}{n(n+1)N^n}\right\},
	\end{equation*}
which are formal power series in $N$, (here $B_n(x)$ denotes the Bernoulli polynomials). This expression follows from the asymptotic expansion of the logarithm of Gamma function. In our context, we only use the property \eqref{gammaconj}, and do not consider the exact expression of the operator \eqref{gammao} in the form of power series.
}

Let $\W_0$ be the space corresponding to the trivial tau-function $1$. We first introduce the following lemma.
\begin{Lem}\label{L1}
	Let $X=\sum_{n=1}^{\infty}a_n\h^nz^{-n}D$, where the set of numbers $\{a_n\}$ are determined by $e^X\cdot z=z\left(1+\frac{\h z^{-1}}{4}\right)^{-1}$. Then, 
	for the space $\mathcal{W}^{\perp}_N$ of the tau-function $\tau_N$ defined by Eq.\eqref{WN}, we have
	\begin{multline*}
		\mathcal{W}^{\perp}_N=\left(1+\frac{\h z^{-1}}{4}(D-N-\frac{1}{2})\right)^{N-\frac{1}{2}}
		\exp\left\{\sum_{n=1}^{\infty}a_n\h^nz^{-n}D\prod_{i=0}^{n-1}\left(D-N-\frac{1}{2}-i\right) \right\}
		\cdot \mathcal{W}_0.
	\end{multline*}
\end{Lem}
\begin{proof}
We first define two operators $U$ and $V$ as
	\begin{align}\label{U}
		U&=1+\sum_{n=1}^{\infty}\frac{(-\h z^{-1})^n}{4^nn!}\prod_{i=0}^{n-1}\left((D-N)^2-(i+\frac{1}{2})^2\right),\\
		V&=1+\sum_{n=1}^{\infty}\frac{(-\h z^{-1})^n}{4^nn!}\prod_{i=0}^{n-1}\left(D-N+i+\frac{1}{2}\right)\nonumber
	\end{align}
The definition of the basis vectors $\Phi_{j}^{(N)}(z)$ in Eq.\eqref{Phi1} implies that $\Phi_{j}^{(N)}(z)=U\cdot z^{j-1}$. Therefore,  $\mathcal{W}^{\perp}_N=U\cdot \mathcal{W}_0$.
Next, using Eq.\eqref{gammaconj}, we can express $U$ as
\begin{align*}
	U&=\sum_{n=0}^{\infty}\frac{(-\h z^{-1})^n}{4^nn!}\frac{\Gamma(D-N+n+\frac{1}{2})}{\Gamma(D-N+\frac{1}{2})}\frac{\Gamma(D-N+\frac{1}{2})}{\Gamma(D-N-n+\frac{1}{2})}\\
	&=\frac{1}{\Gamma(D-N+\frac{1}{2})}\sum_{n=0}^{\infty}\frac{(-\h z^{-1})^n}{4^nn!}\frac{\Gamma(D-N+n+\frac{1}{2})}{\Gamma(D-N+\frac{1}{2})}\Gamma\left(D-N+\frac{1}{2}\right)\\
	&=\frac{1}{\Gamma\left(D-N+\frac{1}{2}\right)}V\Gamma\left(D-N+\frac{1}{2}\right).
\end{align*}
Now, for the operator $V$, we have
\begin{align*}
	V\cdot z^k&=\sum_{n=0}^{\infty}\frac{\h ^n}{4^nn!}\prod_{i=0}^{n-1}\left(-k+N-i-\frac{1}{2}\right)z^{-n+k}\\
	&=(1+\frac{\h z^{-1}}{4})^{N-\frac{1}{2}} \left(z(1+\frac{\h z^{-1}}{4})^{-1}\right)^k.
\end{align*}
The above equation implies that
\begin{equation*}
	V=(1+\frac{\h z^{-1}}{4})^{N-\frac{1}{2}}e^X,
\end{equation*}
where $X$ is the derivation defined in the lemma.
\end{proof}

For the operator $\M_1$ defined by Eq.\eqref{M}, let
\begin{equation}\label{OK}
	\overline{K}=\frac{z}{\h}+\frac{\h}{16}\M_1.
\end{equation}
Consider the space $\overline{\mathcal{W}}^{\perp}=\langle\overline{\Phi}_1(z),\overline{\Phi}_2(z),\dots \rangle$ generated by the basis vectors
\begin{equation*}
	\overline{\Phi}_{j}(z)=\overline{K}^{j-1}\cdot 1,   \mbox{ where } j\geq 1.
\end{equation*}
Then, we have the following result.
\begin{Lem}\label{L2}
For the spaces $\mathcal{W}_N$, $\overline{\mathcal{W}}$ and $\W_0$, we have
\begin{equation}\label{L21}
\mathcal{W}_N=e^{\frac{\h}{4}\M_1^*}\cdot \overline{\mathcal{W}};
\end{equation}
\begin{equation}\label{L22}
\overline{\mathcal{W}}=\exp\left\{-\frac{\h z^{-1}}{4}D\left(D+N+\frac{1}{2}\right)\right\}\operatorname{OE}\left[K_1\right]\cdot \mathcal{W}_0,
\end{equation}
where
\begin{equation}\label{K1}
	K_1=\frac{3 z^{-1}}{4}D\left(D+N+\frac{1}{2}\right)+\frac{\h z^{-2}}{8}(D-1)\left((D+N)^2-\frac{1}{4}\right).
\end{equation}
\end{Lem}
\begin{proof}
For the Kac-Schwarz operator $K_a=\frac{z}{\h}+\frac{1}{2}D$, let us consider the following conjugation
\begin{equation*}
	\overline{K}=e^{\frac{\h}{4}\M_1}K_ae^{-\frac{\h}{4}\M_1}.
\end{equation*}
Using Eq.\eqref{Dcomm}, we have 
\begin{equation*}
	[\M_1,z]=-2D \quad\mbox{and}\quad [\M_1,D]=\M_1.
\end{equation*}
This gives us
\begin{align*}
	& [\frac{\h}{4}\M_1,\frac{z}{\h}+\frac{1}{2}D]=-\frac{1}{2}D+\frac{\h}{8}\M_1,\quad\frac{1}{2!}[\frac{\h}{4}\M_1,[\frac{\h}{4}\M_1,\frac{z}{\h}+\frac{1}{2}D]]=-\frac{\h}{16}\M_1,\\
	& \frac{1}{n!}ad_{\M_1}^n(\frac{z}{\h}+\frac{1}{2}D)=0,\quad n\geq 3.
\end{align*}
Hence, we can obtain the expression of $\overline{K}$ in Eq.\eqref{OK}. Moreover, by Eq.\eqref{phij}, we have
	\begin{equation*}
		\Phi_{1}^{(N)}(z)=e^{-\frac{\h}{4}\M_1}\cdot 1,
	\end{equation*} 
 and we can see from Eq.\eqref{Phij+1} that the following basis vectors
\begin{equation*}
	K_a^{j-1}\cdot \left( e^{-\frac{\h}{4}\M_1}\cdot 1\right)=  e^{-\frac{\h}{4}\M_1}\cdot \left(\overline{K}^{j-1}\cdot 1\right)
\end{equation*}
also generate the space $\mathcal{W}^{\perp}_N$. This implies that $\mathcal{W}^{\perp}_N=e^{-\frac{\h}{4}\M_1}\cdot \overline{\mathcal{W}}^{\perp}$. Using Eq.\eqref{several}, we can obtain Eq.\eqref{L21}.

Next, similar to the proof of Lemma \ref{L1}, for the operator $\M_1$, we have
\begin{align*}
\exp\left(-\frac{\h}{4}\M_1\right)
	=&\exp\left\{\frac{\h z^{-1}}{4}\left((D-\frac{1}{2})^2-N^2\right)\right\}\nonumber\\
	=&\frac{1}{\Gamma\left(D-N+\frac{1}{2}\right)}V_1\Gamma\left(D-N+\frac{1}{2}\right),
\end{align*}
where
\begin{align*}
	V_1=&\exp\left\{\frac{\h z^{-1}}{4}\left(D+N-\frac{1}{2}\right)\right\}
	=\sum_{n=0}^{\infty}\frac{(\h z^{-1})^n}{4^nn!}\frac{\Gamma(D+N+\frac{1}{2})}{\Gamma(D+N-n+\frac{1}{2})}.
\end{align*}
The action of $V_1$ on $z^k$ gives us the following series,
\begin{align*}
	V_1\cdot z^k&=\sum_{n=0}^{\infty}\frac{\h ^n}{4^nn!}\prod_{i=0}^{n-1}\left(k+N-i-\frac{1}{2}\right)z^{-n+k}\nonumber\\
	&=(1+\frac{\h z^{-1}}{4})^{N-\frac{1}{2}} \left(z+\frac{\h }{4}\right)^k.
\end{align*}
Then, we can deduce that
\begin{equation*}
	V_1=(1+\frac{\h z^{-1}}{4})^{N-\frac{1}{2}}e^{\frac{\h }{4}\partial_z}.
\end{equation*}
For the derivation $X$ defined in Lemma \ref{L1}, we let $e^{X_1}=e^{-\frac{\h }{4}\partial_z}e^X$. Then, the operator $U$ in Eq.\eqref{U} can be written as
\begin{align*}
	U=&\frac{1}{\Gamma\left(D-N+\frac{1}{2}\right)}V_1e^{X_1}\Gamma\left(D-N+\frac{1}{2}\right)\\
	=&e^{-\frac{\h}{4}\M_1}\frac{1}{\Gamma\left(D-N+\frac{1}{2}\right)}e^{X_1}\Gamma\left(D-N+\frac{1}{2}\right),
\end{align*}
which means that, for the orthogonal complement space $\overline{\mathcal{W}}^{\perp}$,
\begin{equation*}
	\overline{\mathcal{W}}^{\perp}=\frac{1}{\Gamma\left(D-N+\frac{1}{2}\right)}e^{X_1}\Gamma\left(D-N+\frac{1}{2}\right)\cdot \mathcal{W}_0.
\end{equation*}
For the space $\overline{\mathcal{W}}$, since the adjoint operator of
\begin{equation*}
	\frac{1}{\Gamma\left(D-N+\frac{1}{2}\right)}z^{-k}\Gamma\left(D-N+\frac{1}{2}\right)=z^{-k}\prod_{i=0}^{k-1}\left(D-N-\frac{1}{2}-i\right)
\end{equation*} 
is
\begin{equation*}
	z^{-1}\prod_{i=0}^{k-1}\left(-D-N-\frac{1}{2}-i\right)z^{-k+1}=\Gamma\left(-D-N-\frac{1}{2}\right)z^{-k}\frac{1}{\Gamma\left(-D-N-\frac{1}{2}\right)},
\end{equation*}
by Eq.\eqref{several}, we have
\begin{equation}\label{WperpW0}
	\overline{\mathcal{W}}=\Gamma\left(-D-N-\frac{1}{2}\right)e^{-X_1^*}\frac{1}{\Gamma\left(-D-N-\frac{1}{2}\right)}\cdot \mathcal{W}_0.
\end{equation}

Let us look at the expression of the operator $e^{-X_1^*}$. Suppose $\partial_{\h}e^{X_1}=Ae^{X_1}$, where $A$ is the derivation such that $\operatorname{OE}[A]=e^{X_1}$. From the definition of $X_1$ before, we know that
\begin{equation*}
	e^{X_1}\cdot z=z\left(1-\frac{\h z^{-1}}{4}\right)^2.
\end{equation*}
Then, by comparing the derivatives $\partial_{\h}e^{X_1}\cdot z=Ae^{X_1}\cdot z$, we can obtain the expression of operator $A$ as
\begin{equation*}
	A=-\frac{z^{-1}}{2}\left(1+\frac{\h z^{-1}}{4}\right)^{-1}D.
\end{equation*}
On the other hand, since $\partial_{\h}e^{-X_1}=e^{-X_1}(-A)$, the adjoint operator of $e^{-X_1}(-A)$ would be $(-A^*)e^{-X_1^*}=\partial_{\h}e^{-X_1^*}$. Using Eq.\eqref{adex}, we obtain
\begin{align*}
	-A^*&=-\frac{z^{-1}}{2}D\left(1+\frac{\h z^{-1}}{4}\right)^{-1}=e^{\frac{\h }{4}\partial_z}\left( -\frac{z^{-1}}{2}D+\frac{\h z^{-2}}{8}(D-1)\right)e^{-\frac{\h}{4}\partial_z}.
\end{align*} 
Then, we can compute the following derivative
\begin{align*}
	\partial_{\h}e^{-\frac{\h }{4}\partial_z}e^{-X_1^*}&=e^{-\frac{\h }{4}\partial_z}\left(-\frac{z^{-1}}{4}D-A^*\right)e^{-X_1^*}\\
	&=\left(-\frac{3z^{-1}}{4}D+\frac{\h z^{-2}}{8}(D-1) \right)e^{-\frac{\h }{4}\partial_z}e^{-X_1^*}.
\end{align*}
This implies that
\begin{equation}\label{X14}
	e^{-X_1^*}=e^{\frac{\h }{4}\partial_z}\operatorname{OE}\left[-\frac{3z^{-1}}{4}D+\frac{\h z^{-2}}{8}(D-1) \right].
\end{equation}
Finally, Eq.\eqref{L22} follows from Eq.\eqref{WperpW0} and \eqref{X14}.
\end{proof}

Let $\overline{\tau}$ be the tau-function determined by the space $\overline{\mathcal{W}}$. Note that the energy of the operator in Eq.\eqref{M1} is positive. Hence, by Eq.\eqref{WG}, \eqref{several} and \eqref{L21}, we have
\begin{equation}\label{tauNM1}
	\tau_N(q)=\exp\left\{\frac{\h}{4}\widehat{M}_{-1}+\frac{\h}{4}\widehat{L}_{-1}+\left(\frac{\h}{16}-\frac{\h N^2}{4}\right)q_1 \right\} \cdot\overline{\tau}(q).
\end{equation}
Moreover, let $\overline{\tau}_1$ be the tau-function determined by the space $\operatorname{OE}\left[K_1\right]\cdot \mathcal{W}_0$, where $K_1$ is defined in Eq.\eqref{K1}. Then, by Eq.\eqref{L22} we have
\begin{equation}\label{tau1}
	\overline{\tau}(q)=\exp\left\{\frac{\h}{4}\widehat{M}_{-1}+\left(\frac{\h N}{4}+\frac{\h}{8}\right)\widehat{L}_{-1}\right\}\cdot\overline{\tau}_1(q).
\end{equation}
Finally, the tau-function $\overline{\tau}_1(q)$ has the following representation.
\begin{Prop}
	For the operator $K_1$ defined in Eq.\eqref{K1}, we have
\begin{equation*}
	\overline{\tau}_1=\operatorname{OE}[\widehat{W}_{K_1}]\cdot 1,
\end{equation*}
where
\begin{equation}\label{P4}
	\widehat{W}_{K_1}=-\frac{3}{4}\widehat{M}_{-1}-\frac{6N+3}{8}\widehat{L}_{-1}-\frac{\h}{8}\widehat{Q}_{-2}-\frac{\h(4N+1)}{16}\widehat{M}_{-2}- \frac{\h(4N^2-1)}{32}\left(\widehat{L}_{-2}-\frac{1}{2}q_2\right)
\end{equation}
\end{Prop}
\begin{proof}
The vectors $\{\Phi_{j}^{(1)}(z)=\operatorname{OE}\left[K_1\right]\cdot z^{j-1}\}$ form a set of basis vectors for the tau-function $\overline{\tau}_1$. Furthermore, we can see that the equation 
$$D_\h\cdot \Phi_{j}^{(1)}(z)=\h K_1\cdot \Phi_{j}^{(1)}(z)=(D_z-j+1)\cdot \Phi_{j}^{(1)}(z)$$ 
uniquely determines all those basis vectors. In other words, $D_z-\h K_1$ is a Kac-Schwarz operator that uniquely determines the point of the Sato Grassmannian for $\overline{\tau}_1$. This is equivalent to say that the equation
\begin{equation*}
	\left(\widehat{L}_0-\h\widehat{W}_{ K_1}\right)\overline{\tau}_1(q)=\left(D_{\h}-\h\widehat{W}_{ K_1}\right)\overline{\tau}_1(q)=C\h\overline{\tau}_1(q)
\end{equation*}
uniquely determines the tau-function up-to a constant $C$. (In fact, this gives us a $W_{1+\infty}$ constraint that determines $\overline{\tau}_1$.) Since $\widehat{W}_{K_1}$ is a $W_{1+\infty}$ operator with positive energy, we can see that $\operatorname{OE}[\widehat{W}_{K_1}]\cdot 1$ is a KP tau-function that satisfies
\begin{equation}\label{EqC}
	\left(\partial_{\h}-\widehat{W}_{ K_1}-C\right)\cdot\left(\operatorname{OE}[\widehat{W}_{K_1}]\cdot 1\right)=0.
\end{equation}
Using Eq.\eqref{nodes1}, \eqref{nodes2} and \eqref{nodes3}, we have the following operator expressions 
\begin{align*}
	&W_{-z^{-2}D}=L_{-2}+\frac{1}{2}\alpha_{-2},\quad W_{-z^{-2}D^2}=M_{-2}+L_{-2}+\frac{1}{2}\alpha_{-2},\\
	&W_{-z^{-2}D^3}=Q_{-2}+\frac{3}{2}M_{-2}+L_{-2}+\frac{1}{2}\alpha_{-2}.
\end{align*}
This will give us the expression of $\widehat{W}_{K_1}$ in Eq.\eqref{P4}, and it also tells us that the constant $C$ in Eq.\eqref{EqC} should be $0$. Therefore,   $\overline{\tau}_1(q)=\operatorname{OE}[\widehat{W}_{K_1}]\cdot 1$. 
\end{proof}

Using Eq.\eqref{tauNM1}, \eqref{tau1} and \eqref{P4}, we have completed the proof of Theorem \ref{main} in this subsection.

\vspace{6pt}
\noindent
{\bf Remark}: 
\emph{By taking the derivative $\partial_{\h}$ on Eq.\eqref{maineq}, we can obtain a $W_{1+\infty}$ constraint for $\tau_N$. Equivalently, $\tau_N$ can be expressed as the ordered exponential of this constraint acting on $1$.
}

\subsection{The BGW tau-function $\tau_{BGW}$}\label{S32}
Let $\overline{M}_{g,n}$ be the moduli space of complex stable curves of genus $g$ with $n$ marked points, and $\bpsi_i$ be the first Chern class of the cotangent line over $\overline{M}_{g,n}$ at the $i$th marked point. The BGW tau-function $\tau_{BGW}$ is conjectured to be governing the intersection numbers 
\begin{equation*}
	\int_{\overline{M}_{g,n}}\Theta_{g,n}\bpsi_1^{d_1}\dots \bpsi_n^{d_n},
\end{equation*}
where $\Theta_{g,n}$ denotes the cohomology class constructed via the push-forward map using the moduli space of stable twisted spin curves, (\cite{NP}, \cite{NP1}). It is defined to be zero when the numbers $d_i$ and genus $g$ do not satisfy the condition 	$\sum_{i=1}^n d_i=g-1$. 

We define the generating function $F_{BGW}(q)$ to be
\begin{equation*}
	F_{BGW}(q)=\sum_{g,n\geq 0}\frac{1}{n!}\sum_{d_1,\dots,d_n\geq 0} \int_{\overline{M}_{g,n}}\Theta_{g,n}\bpsi_1^{d_1}\dots \bpsi_n^{d_n} \prod_{i=1}^n\frac{q_{2d_i+1}}{2d_i+1}.
\end{equation*}
Then the tau-function of $Z_{BGW}$ is conjectured to be $\tau_{BGW}=\exp(F_{BGW}(q))$ under the variable change $q_{k}=\tr \Lambda^{-k}$ \cite{NP}. The determinantal representation of $Z_{BGW}$ in the Kontsevich phase has been introduced in \cite{A}, that is,
\begin{equation}
	Z_{BGW}=\frac{\det_{i,j=1}^M\Phi^{(0)}_j(z_i)}{\Delta(z)},
\end{equation}
where $\{\Phi^{(0)}_j\}$ are the basis vectors in the form
\begin{equation*}
	\Phi^{(0)}_j(z)=\sqrt{4\pi z}z^{j-1}e^{-2z}I_{j-1}(2z),
\end{equation*}
and $I_{\nu}(z)$ is the modified Bessel function. The asymptotic expansion of $\Phi^{(0)}_j(z)$ for large values of $|z|$ with $\arg(z)\neq \pi$  corresponds to the case $N=0$ in Eq.\eqref{Phi1}.

\subsection{The tau-function of Grothendieck’s dessins d'enfant}\label{S33}
A Belyi pair $(C,f)$ consists of a smooth algebraic curve over $\C$ with a meromorphic function $f$ on $C$ ramified at three points $\{0,1,\infty\}$. Let $g$ be the genus of $C$ and $d$ be the degree of $f$. Let $k,l$ and $m$ be the cardinality of the pre-image of $0,1$ and $\infty$ respectively. Then, by Riemann-Hurwitz formula, we have $2g-2=d-(k+l+m)$. There is a one-to-one correspondence between the isomorphism classes of Belyi pairs and connected bicolored ribbon graphs known as Grothendieck’s dessins d'enfant \cite{G}.

Suppose $\mu=(\mu_1,\dots,\mu_m)$ is a partition of the degree $d$, corresponding to the orders of the $m$ poles of $f$. The number $N_{k,l}(\mu)$ denotes the weighted count of labeled dessins d'enfant of the type $(k,l,\mu)$. The tau-function $\tau_D$ of such dessins d'enfant is defined by $\tau_D=\exp(F_D)$, where
\begin{equation*}
 F_D=\sum_{k,l,m\geq 1}\frac{1}{m!}\sum_{\mu_1,\dotsm\mu_m\geq 1}N_{k,l}(\mu_1,\dots,\mu_m)s^du^kv^lq_{\mu_1}\dots q_{\mu_m}.
\end{equation*}
It can be represented as a $W_{1+\infty}$ operator acting on the trivial tau-function $1$, (\cite{MZo}, \cite{PZ}). 
\begin{equation}\label{dessins}
	\tau_D(q)=\exp\left\{s\left(\widehat{M}_{-1}+(u+v)\widehat{L}_{-1}+uvq_1\right) \right\}\cdot 1.
\end{equation}
The operator in Eq.\eqref{tauNM1} is the case $s=\frac{\h}{4}, u=\frac{1}{2}+N,v=\frac{1}{2}-N$ of the above operator.

\subsection{The Kontsevich-Witten tau-function}\label{S34}
The intersection numbers of the $\bar{\psi}$-classes are evaluated by the integral:
$$\left<\btau_{d_1}\dots \btau_{d_n}\right>=\int_{\overline{M}_{g,n}}\bpsi_1^{d_1}\dots \bpsi_n^{d_n}.$$
It is defined to be zero when the numbers $d_i$, $n$ and genus $g$ do not satisfy the condition 
\begin{equation}\label{condition}
	\sum_{i=1}^n d_i=\dim(\overline{M}_{g,n})=3g-3+n.
\end{equation}
We define the generating function $F_K(q)$ to be
\begin{equation*}
	F_K(q)=\left< \exp\left( \sum_{k=0}^{\infty} (2k-1)!!q_{2k+1}\btau_{k}\right)\right>.
\end{equation*}
The Kontsevich matrix model is described by the following integral over the space of Hermitian matrices
\begin{equation*}
	Z_{KW}=\frac{\int [\d\Phi] \exp\left(-\tr (\frac{\Phi^3}{6}+\frac{\Lambda\Phi^2}{2}) \right)}{\int [\d\Phi] \exp\left(-\tr \frac{\Lambda\Phi^2}{2} \right)},
\end{equation*}
where $\Lambda$ is the diagonal matrix. It gives us a representation of the function $F_K(q)$ with
$$q_{2k+1}=\tr(\Lambda^{-2k-1}),$$
that is, $\tau_{KW}(q)=\exp(F_K(q))$. It is well-known that $\tau_{KW}$ is a tau-function for the KdV hierarchy (\cite{K}). An explicit formula of the Fermionic representation for $Z_{KW}$ can be found in \cite{Z1} (see also \cite{BY}).  

The determinantal representation of $Z_{KW}$ has been introduced in \cite{AE}, that is,
\begin{equation*}
	Z_{KW}=\frac{\det_{i,j=1}^M\Phi^{(KW)}_j(z_i)}{\Delta(z)},
\end{equation*}
where the basis vectors in $\mathcal{W}_{KW}=\{\Phi^{(KW)}_1,\Phi^{(KW)}_2,\dots \}$ are determined by the integrals
\begin{equation*}
	\Phi^{(KW)}_j(z)=\frac{z^{\frac{1}{2}}e^{-\frac{z^3}{3}}}{\sqrt{2\pi }}\int_{-\infty}^{\infty} y^{j-1}\exp\left(-\frac{y^3}{6}+\frac{yz^2}{2}\right)\mathrm{d} y,
\end{equation*}
and they can be generated by $\Phi_1^{(KW)}(z)$ using a Kac-Schwarz operator as follows,
\begin{align}
	&\Phi_1^{(KW)}(z)=\sum_{k=0}^{\infty}\frac{2^k\Gamma(3k+\frac{1}{2})}{9^k(2k!)\Gamma(\frac{1}{2})}z^{-3k};\label{phi1K}\\
	&\Phi^{(KW)}_{j}(z)=\left(z +z^{-1}\partial_z-\frac{ z^{-2}}{2}\right)^{j-1}\cdot \Phi_1^{(KW)}(z)\label{KSw}.
\end{align}

For the Kontsevich-Witten tau-function $\tau_{KW}$, we give a similar result as Eq.\eqref{tauNM1}. 
\begin{Prop}
	Let $\widetilde{\W}^{\perp}$ be the space generated by the basis vectors $\widetilde{K}^{j-1}\cdot 1$, where
	\begin{equation*}
		\widetilde{K}=z^{-2}\left(1-\frac{z^{-3}}{2}(D-\frac{5}{2})\right)^{-\frac{4}{3}}\left(z^3-\frac{1}{4}z^{-3}(D-\frac{1}{2})(D-\frac{5}{2})-\frac{1}{2}\right).
	\end{equation*}
	and let $\widetilde{\tau}$ be the tau-function corresponding to the space $\widetilde{\W}^{\perp}$. Then, 
	\begin{equation}\label{tauKM3}
		\tau_{KW}(q)=  \exp\left(\frac{1}{6}\widehat{M}_{-3}+\frac{1}{6}\widehat{L}_{-3}-\frac{1}{72}q_{3} \right)\cdot \widetilde{\tau}(q)
	\end{equation}	
\end{Prop}
\begin{proof}
The proof is similar to the one of Lemma \ref{L2}. Notice that we can express the first basis vector $\Phi_1^{(KW)}(z)$ in Eq.\eqref{phi1K} as follows:
\begin{align*}
	\Phi_1^{(KW)}(z)
	&=\sum_{k=0}^{\infty}\frac{6^k\Gamma(k+\frac{1}{2})}{\Gamma(\frac{1}{2})(2k!)}\frac{\Gamma(k+\frac{1}{6})}{\Gamma(\frac{1}{6})}\frac{\Gamma(k+\frac{5}{6})}{\Gamma(\frac{5}{6})}z^{-3k}\\
	&=\sum_{k=0}^{\infty}\frac{3^k}{2^kk!}\prod_{i=1}^{k}(-\frac{1}{3}D+\frac{1}{6}-i)(-\frac{1}{3}D+\frac{5}{6}-i)z^{-3k}\cdot 1\\
	&=\sum_{k=0}^{\infty}\frac{1}{6^kk!}\left(z^{-3}(D-\frac{1}{2})(D-\frac{5}{2}) \right)^k\cdot 1\\
	&=\exp\left( \frac{1}{6}\M_3^{(K)}\right)\cdot 1,
\end{align*}
where 
\begin{equation*}
	\M_3^{(K)}=z^{-3}(D-\frac{1}{2})(D-\frac{5}{2}).
\end{equation*}
Let us write the Kac-Schwarz operator in Eq.\eqref{KSw} as $z^{-2}(z^3+D-\frac{1}{2})$.  Using Eq.\eqref{Dcomm}, we have
\begin{equation*}
	[\M_3^{(K)},z^3]= 6D \quad\mbox{and}\quad [\M_3^{(K)},D]=3\M_3^{(K)}.
\end{equation*}
Then, we compute the following two conjugations,
\begin{align*}
 &\exp\left( -\frac{1}{6}\M_3^{(K)}\right)(z^3+D-\frac{1}{2}) \exp\left( \frac{1}{6}\M_3^{(K)}\right)
 =z^3-\frac{1}{4}\M_3^{(K)}-\frac{1}{2},\\
	&\exp\left( -\frac{1}{6}\M_3^{(K)}\right)z^{-2} \exp\left( \frac{1}{6}\M_3^{(K)}\right)\\
	=&z^{-2}\left(1+\sum_{n=1}^{\infty} \frac{1}{6^nn!}\prod_{i=0}^{n-1}(4+3i)z^{-3n}\prod_{i=0}^{n-1}(D-\frac{5}{2}-3i)\right)\\
	=&z^{-2}\left\{1+\sum_{n=1}^{\infty} \frac{(-1)^n}{2^nn!}\frac{\Gamma(-\frac{1}{3})}{\Gamma(-\frac{1}{3}-n)}\left(z^{-3}(D-\frac{5}{2}) \right)^n\right\}\\
	=&z^{-2}\left(1-\frac{z^{-3}}{2}(D-\frac{5}{2})\right)^{-\frac{4}{3}}.
\end{align*}
This gives us the expression of the Kac-Schwarz operator $\widetilde{K}$. The ajoint operator of $\M_3^{(K)}$ is
\begin{equation}\label{M3}
	z^{-1}(D+\frac{1}{2})(D+\frac{5}{2})z^{-2}=z^{-3}D^2-z^{-3}D-\frac{3}{4}z^{-3}.
\end{equation}
From Eq.\eqref{nodes1} and \eqref{nodes2}, we have
\begin{equation*}
	W_{-z^{-3}D}=L_{-3}+\alpha_{-3}, \quad W_{-z^{-3}D^2}=M_{-3}+2L_{-3}+\frac{5}{3}\alpha_{-3}.
\end{equation*}
Therefore the operator \eqref{M3} corresponds to the fermionic operator $-M_{-3}-L_{-3}+\frac{1}{12}\alpha_{-3}$ with positive energy. 
\end{proof}
	\vspace{10pt}
\noindent
{\bf Remark:}
\emph{
 Let us consider the following basis vectors for the Kontsevich-Penner model (see, e.g., \cite{AO1},\cite{AO2} and references therein):
\begin{equation*}
	\Phi^{(KP)}_j(z)=\frac{z^{N+\frac{1}{2}}e^{-\frac{z^3}{3}}}{\sqrt{2\pi }}\int_{-\infty}^{\infty} y^{j-1-N}\exp\left(-\frac{y^3}{6}+\frac{yz^2}{2}\right)\mathrm{d} y.
\end{equation*}
When $N=0$ they are the basis vectors $\Phi^{(KW)}_j(z)$ for $\tau_{KW}$. If we express them in terms of the power series:
\begin{multline*}
	\Phi^{(KP)}_{j}(z)=z^{j-1}\sum_{n=0}^{\infty}\binom{j-1-N}{2n}\sum_{k=0}^{\infty}\frac{2^{k+n}\Gamma(3k+n+\frac{1}{2})}{3^{2k}(2k)!\Gamma(\frac{1}{2})}z^{-3k-3n}\\
	\quad-z^{j-1}\sum_{n=0}^{\infty}\binom{j-1-N}{2n+1}\sum_{k=0}^{\infty}\frac{2^{k+n+1}\Gamma(3k+n+\frac{5}{2})}{3^{2k+1}(2k+1)!\Gamma(\frac{1}{2})}z^{-3k-3n-3}.
\end{multline*}
we can obtain a $W_{1+\infty}$ representation for the Kontsevich-Penner model using the operator 
\begin{align*}
	&\sum_{k=0}^{\infty}\frac{2^k\Gamma(3k+\frac{1}{2})}{3^{2k}(2k)!\Gamma(\frac{1}{2})}z^{-3k}\\
	&+\sum_{k=0,n=1}^{\infty}\frac{2^{k+n}\Gamma(3k+n+\frac{1}{2})}{3^{2k}(2n)!(2k)!\Gamma(\frac{1}{2})}z^{-3k-3n}\prod_{i=1}^{2n}(N-D-1+i)\\
	&-\sum_{k,n=0}^{\infty}\frac{2^{k+n+1}\Gamma(3k+n+\frac{5}{2})}{3^{2k+1}(2n+1)!(2k+1)!\Gamma(\frac{1}{2})}z^{-3k-3n-3}\prod_{i=1}^{2n+1}(N-D-1+i).
\end{align*}
However, unlike the operator $U$ in Eq.\eqref{U}, this one seems a little complicated to be decomposed into several simple ones.
}

\subsection{Linear Hodge tau-function}\label{S35}
Let $\lambda_j$ be the $j$th Chern class of the Hodge bundle over $\overline{M}_{g,n}$. The linear Hodge integrals are the intersection numbers of the form
$$\left<\lambda_j\tau_{d_1}\dots \tau_{d_n}\right>=\int_{\overline{M}_{g,n}}\lambda_j\bpsi_1^{d_1}\dots \bpsi_n^{d_n}.$$
They are defined to be zero when the numbers $j$ and $d_i$ do not satisfy the condition
\begin{equation*}
	j+\sum_{i=1}^n d_i=\dim(\overline{M}_{g,n})=3g-3+n.
\end{equation*} Let $\widetilde{\phi_0}(u,q)=q_1$, and for $k\geq 0$,
\begin{equation}\label{def:phi}
	\widetilde{\phi_{k+1}}(u,q)=\left( \widehat{L}_{-2}+2u\widehat{L}_{-1}+u^2\widehat{L}_0-\frac{1}{2}q_1^2 \right)\cdot \widetilde{\phi_k}(u,q).
\end{equation}
The function $F_{Ho}(u,q)$ is defined as
\begin{equation}\label{FHuq}
	F_{Ho}(u,q)=\sum (-1)^j\left<\lambda_j\tau_{d_1}\dots \tau_{d_n}\right>u^{2j}\prod_{i=1}^n \widetilde{\phi_{d_i}}(u,q),
\end{equation}
and $\tau_{Ho}(q)=\exp(F_{Ho}(u,q))$. The matrix model for $Z_{Ho}$ is introduced in \cite{AE}. If we set $u=0$, the polynomial $\widetilde{\phi_k}(u,q)$ only has the term $(2k-1)!!q_{2k+1}$, and $\exp(F_{Ho}(0,q))$ is exactly the Kontsevich-Witten tau-function $\tau_{KW}(q)$.

There are several formulas connecting $Z_{KW}$ to $Z_{Ho}$ using the $W_{1+\infty}$ operators, and they can be transformed to each other through some computations. From the results in \cite{AE} and \cite{W}, we have
\begin{align}\label{Aformula}
	\tau_{KW}(q) &= \exp\left(\sum_{k>0}\widehat{a}_k^{(1)}u^k\widehat{L}_k\right)\exp(\widehat{N})\cdot \tau_{Ho}(q)\nonumber\\
	&= \exp\left(\sum_{k>0}\widehat{a}_k^{(1)}u^k\widehat{L}_k\right)\exp\left(\sum_{k>0}\widehat{a}_k^{(2)}u^k\widehat{L}_k\right)\cdot \tau_{Ho}(q)\nonumber\\
	&= \exp\left(\sum_{k>0}\widehat{a}_ku^k\widehat{L}_k\right)\cdot \tau_{Ho}(q),
\end{align}
where the coefficients $\{\widehat{a}_k^{(i)}\}$ and $\{\widehat{a}_k\}$ are determined by the series
\begin{equation*}
	f_{+} =\exp(\sum_{k=1}^{\infty} \widehat{a}_kz^{1-k}\frac{\partial}{\partial z})\cdot z,\quad
	f_{+i} = \exp\left(\sum_{k>0}\widehat{a}_k^{(i)}z^{1+k}\frac{\d}{\d z}\right)\cdot z,
\end{equation*}
with $f_{+}(z) = f_{+1}(f_{+2}(z))$,
\begin{align*}
	& f_{+1} = \frac{1}{z\exp(z^{-1})\sinh(z^{-1})-1},\\
	& \frac{1}{(f_{+2})^2}\coth\left(\frac{1}{f_{+2}}\right)-\frac{1}{f_{+2}}=\frac{1}{3z^3},
\end{align*}
and, for Bernoulli numbers $B_{2k}$,
\begin{equation*}
	\widehat{N}=\sum_{k\geq 2}\frac{2^{2k}B_{2k}}{(2k)!}u^{2k-2}\frac{(2k+1)\partial}{\partial q_{2k+1}}.
\end{equation*}
Furthermore, from the results in \cite{LW}, we also have
\begin{align}\label{LWformula}
	\tau_{Ho}(q) &= \exp(\sum_{m>0} a_mu^m\widehat{L}_m)\exp(P)\cdot \tau_{KW}(q),\nonumber\\
	&= \exp\left(\sum_{m=1}^{\infty} e_mu^m\widehat{L}_m\right)\cdot \tau_{KW}(q),
\end{align}
where
\begin{align*}
	&\exp(\sum_{m>0}a_mz^{1-m}\frac{\partial}{\partial z})\cdot z=(-2\ln(1-\frac{1}{1+z})-\frac{2}{1+z})^{-\frac{1}{2}},\\
	&\exp(\sum_{m=1}^{\infty} e_{m}z^{1-m}\frac{\partial}{\partial z})\cdot z=\left(3\sum_{k=0}^{\infty} \frac{b_{2k+1}}{2k+3}z^{-2k-3}\right)^{-\frac{1}{3}}
\end{align*}
and
\begin{equation*}
	P = -\sum_{k=1}^{\infty} b_{2k+1}u^{2k}\frac{\partial}{\partial q_{2k+3}},
\end{equation*}
with $b_1=1, b_2=1/3$,
\begin{equation*}
	(n+1)b_n=b_{n-1}-\sum_{k=2}^{n-1}kb_kb_{n+1-k}.
\end{equation*}
Note that $-e_k=\widehat{a}_k$ for $k\geq 1$. 

\subsection{Hurwitz tau-function}\label{S36}
Let $H$ be the generating function of simple Hurwitz numbers, (for details about the simple Hurwitz numbers, we refer the readers to \cite{OP} and references therein). The function $\exp(H)$ has a cut-and-join representation \cite{GJ}.
\begin{equation}\label{cutandjoin}
	e^H=e^{\frac{\beta}{2} \widehat{M}_0} e^{q_1}\cdot 1,
\end{equation}
and this representation indicates that the function $\tau_H=\exp(H)$ is a tau-function for the KP hierarchy. In \cite{MK}, Kazarian obtained the function $F_{Ho}(u,q)$ from the Hurwitz generating function $H$ using the ELSV formula (see \cite{ELSV}). In fact, setting $u=\beta^{\frac{1}{3}}$, we have (Theorem 2.3 in \cite{MK})
\begin{equation}\label{Ke}
	\exp(-\sum_{m=1}^{\infty} a_{-m}\beta^m\sum_{i=1}^{\infty} iq_{i+m}\frac{\partial}{\partial q_i})
	\cdot (H-H_{0,1}-H_{0,2})=\exp(\frac{4}{3}\ln{\beta} L_0)\cdot F_{Ho}(\beta^{\frac{1}{3}},q),
\end{equation}
where
$$H_{0,1}=\sum_{b=0}^{\infty}\frac{b^{b-2}}{b!}\beta^{b-1}q_b, \quad H_{0,2}=\frac{1}{2}\sum_{b_1,b_2=1}^{\infty}\frac{b_1^{b_1}b_2^{b_2}}{(b_1+b_2)b_1!b_2!}\beta^{b_1+b_2}q_{b_1}q_{b_2}.$$
And the numbers $\{a_{-m}\}, m>0,$ are determined by
\begin{equation*}
	\exp(-\sum_{m=1}^{\infty} a_{-m}z^{1+m}\frac{\partial}{\partial z})\cdot z=\frac{z}{1+z}e^{-\frac{z}{1+z}}.
\end{equation*}
If we let
\begin{equation*}
	F_{Ho}(\beta)=\exp(\frac{4}{3}\ln\beta \widehat{L}_0)\cdot F_{Ho}(\beta^{\frac{1}{3}},q),
\end{equation*}
Then Eq.\eqref{Ke} is equivalent to (see also \cite{AE},\cite{GW})
\begin{equation}\label{Kformula}
	\exp(H)=\exp(H_{0,1})\exp(\sum_{m>0} a_{-m}\beta^m\widehat{L}_{-m})\cdot \exp(F_{Ho}(\beta)),
\end{equation}
and this formula also preserves the KP integrability.

\section*{Acknowledgement}
The author would like to thank Alexander Alexandrov and Shuai Guo for helpful discussions. Research of the author is supported by the National Natural Science Foundation of China (Grant No.11701587).

\section*{Appendix}
\subsection*{A.1 Magnus expansion}
The ordered exponential is the unique solution of the differential equation:
\begin{equation*}
	\frac{\partial}{\partial \h} \operatorname{OE}[A](\h) = A(\h) \operatorname{OE}[A](\h)
	\quad\mbox{and}\quad\operatorname{OE}[A](0) = 1.
\end{equation*}
It can be represented as the infinite summation:
\begin{equation*}
	\operatorname{OE}[A](\h) = 1 + \int_0^{\h} A(\h_1) \, \d\h_1+ \int_0^\h \int_0^{\h_1} A(\h_1) A(\h_2) \, \d\h_2 \, \d\h_1 + \cdots.
\end{equation*}
Let $ \operatorname{OE}[A](\h)=\exp\left(\Omega(\h)\right)$. Then
\begin{equation*}
 \partial_\h\Omega(\h)=\frac{ad_{\Omega}}{\exp\left(ad_{\Omega}\right)-1}A=\sum_{n=0}^{\infty}\frac{B_n}{n!}ad_{\Omega}^{n}A.
\end{equation*}
This provides a recursive procedure to compute all the terms in the Magnus expansion: $\Omega(\h)=\sum_{n=1}^{\infty}\Omega_n$. Starting from $\Omega_1 = \int_0^{\h} A(x) \, \d x$, we have
\begin{equation*}
\Omega_n = \sum_{j=1}^{n-1} \frac{B_j}{j!} \int_0^\h S_n^{(j)}(x) \, \d x , \quad n \geq 2,
\end{equation*}
where
\begin{equation*}
	S_n^{(j)} = \sum_{m=1}^{n-j} \left[\Omega_m, S_{n-m}^{(j-1)}\right], \quad 2 \leq j \leq n - 1,
\end{equation*}
\begin{equation*}
	S_n^{(1)} = \left[\Omega_{n-1}, A\right], \quad S_n^{(n-1)} = \operatorname{ad}_{\Omega_1}^{n-1}(A).
\end{equation*}
This procedure is known as Magnus generator.

\vspace{10pt} \noindent
\\
\footnotesize{\sc gehao wang }\\
Department of Mathematics,\\
College of Information Science and Technology/College of Cyberspace Security,\\
Jinan University, Guangzhou, China. \\
\footnotesize{E-mail address:  gehao\_wang@hotmail.com}

\end{document}